\documentclass[reprint,
superscriptaddress,
 amsmath,amssymb,
 aps,
floatfix,
]{revtex4-2}

\usepackage{times} %Times new roman
\usepackage{graphicx}% Include figure files
\usepackage{dcolumn}% Align table columns on decimal point
\usepackage{bm}% bold math
\usepackage{hyperref}% add hypertext capabilities
\begin{document}

\preprint{APS/123-QED}

\title{Generating broadband optical squeezing via Cascaded Micro-Ring Resonators}% Force line breaks with \\
%% \thanks{A footnote to the article title}%

\author{Chung-Hsien Wang}
\email{chwang22195@uchicago.edu}
\affiliation{Department of Physics, The University of Chicago,
Chicago, Illinois 60637, USA}
\author{Tian Zhong}
\email{tzh@uchicago.edu}
\affiliation{Pritzker School of Molecular Engineering, University of Chicago, Chicago, Illinois 60637, USA}

\date{\today}% It is always \today, today,

\begin{abstract}
Broadband squeezed light functioning as a Markovian reservoir can exponentially enhance light-matter interactions, benefiting quantum technologies. However, conventional single-cavity sources face a trade-off between squeezing depth and spectral bandwidth. We propose a scalable scheme for generating broadband squeezed vacuum using a cascade of parametric microring resonators coupled to a common bus waveguide. By analyzing the output, we identify the specific  conditions that yield a broad, flat-topped squeezing spectrum, even under realistic intracavity pump attenuation. We demonstrate that this architecture is robust against fabrication imperfections, including inhomogeneous resonator frequencies and component failures. We show that the flat-topped spectrum converges to the Markovian limit significantly faster than a single-cavity Lorentzian profile. An array of as few as $N=5$ coupled resonators with an intrinsic loss ratio of $\kappa_I/\kappa = 0.1$ reduces the required bandwidth to a quarter of that needed by a single cavity to achieve same squeezing. This rapid convergence relaxes the low-$Q$ and high-gain constraints of single broadband cavities, distributing the squeezing process across moderately pumped resonators to provide a practical route for engineering squeezed reservoirs on mature integrated photonic platforms.
\end{abstract}

\maketitle

%\tableofcontents

\section{Introduction}
Squeezed states of light are a crucial resource for continuous-variable (CV) quantum information and metrology. One of its well-known applications is the gravitational-wave detections \cite{Aasi2013Squeezing,PhysRevLett.123.231108}, where shrunk quadrature carries high precisions to quantum imaging and sensing. Squeezing is also combined with Gaussian boson sampling to realize programmable processors \cite{madsen2022quantum}, and powers engineered dissipative reservoirs that stabilize entangled steady states across atomic, superconducting, and optomechanical platforms \cite{PhysRevLett.107.080503,Murch2013ObservingSQ,Wollman2015}. In these settings, practical advantage relies on broad spectral support rather than merely peak squeezing magnitude. This broadband capability is necessary, for example, to multiplex large time-domain cluster states for CV computation \cite{Larsen2019Deterministic} or to suppress radiative decay in atomic systems \cite{murch2013reduction}. Consequently, developing genuinely broadband squeezed sources is essential to realize the high processing bandwidths and robust entanglement fidelities demanded by next-generation CV protocols.

%%%%%%%%%%%%%%%%%%%%%%%%%%%%%%

Existing routes to optical squeezing face a trade-off between peak amplitude and spectral flatness. While doubly resonant optical parametric amplifiers can achieve remarkable squeezing exceeding 15 dB \cite{PhysRevLett.117.110801}, they are restricted to MHz-scale linewidths. By contrast, integrated microring platforms based on silicon nitride, AlGaAs, and GeSbSe can deliver multi-GHz on-chip squeezing \cite{vaidya2020broadband,PRXQuantum.2.010337,Zhang2021HighQ,PhysRevApplied.3.044005}, with a more moderate noise reduction.  At the opposite side, single-pass nonlinear waveguides can deliver squeezing over terahertz bandwidths \cite{10.1063/1.5142437,Ledezma:22}. Without resonant field enhancement, however, the same parametric gain must be supplied by higher pump intensities. Bridging this gap is crucial, as recent work demonstrates that a broadband squeezed source can generate an effective Markovian reservoir \cite{8qtt-symt}, that exponentially enhances light–matter interactions \cite{PhysRevLett.120.093602,PhysRevLett.120.093601,Lemonde2016}. This mechanism provides a pathway to ultrastrong-coupling regimes that is difficult with an vacuum bath, benefiting protocols such as fast quantum gates \cite{PhysRevResearch.3.033275} and in-situ quantum memories \cite{PhysRevLett.134.053602} that crave for high coupling strengths. To implement this scenario, however, the external source spectrum must maintain uniform  across the target system's linewidth, ensuring that the bath appears Markovian on the timescale of the system dynamics. This flatness imposes strict parameter constraints, particularly because all single-cavity resonant methods naturally yield a Lorentzian profile. Therefore, we require architectures beyond standard single-cavity geometries to overcome these  constraints.

%%%%%%%%%%%%%%%%%%%%%%%%%%%%%%%%%

In this work, we propose a scheme that meets this requirement using cascaded micro-ring resonators coupled to a common waveguide bus. Cascaded cavities have been utilized in various frameworks, such as enhancing the indistinguishability of photons \cite{PhysRevLett.122.183602} and realizing high-transmission diodes \cite{li2017cascaded}, and have recently been shown to produce broadband sources in chiral cavity-magnonic systems \cite{PhysRevA.111.023701}.  A related line of theoretical work has studied cascaded optical parametric amplifiers \cite{PhysRevA.87.023834} and delayed-feedback schemes \cite{PhysRevA.94.023809}, showing that both routes enhance squeezing beyond a single stage. These analyses, however, target the peak squeezing of identical resonators rather than the spectral shape and disorder robustness that determine whether a lossy, non-identical integrated chain can act as a Markovian reservoir. In our proposed picture, each ring acts as a low-Q parametric source whose individual squeezing spectrum is Lorentzian. Nevertheless, the feed-forward coupling causes the noise quadratures to be transmitted multiplicatively along the chain. As a result, the output spectrum becomes the $N$-th power of a single-cavity transfer function, causing both the depth and the width of the squeezing dip to grow with the number of rings. Accounting for realistic intrinsic loss and the geometric attenuation of the intracavity pump, which causes the per-ring pair-generation rate to decay along the chain, we derive a closed-form expression for the steady-state squeezing spectrum. This provides the flexibility to engineer a flat spectrum by selecting a pumping power that balances the broadened Lorentzian profile against the sag caused by pump attenuation. Furthermore, we demonstrate the robustness of this scheme against experimental imperfections, such as inhomogeneous resonator frequencies or defective rings within the chain. Finally, by coupling the resulting non-Lorentzian noise correlations to an arbitrary target system, we use a non-Markovianity measure to quantify how rapidly the cascade approaches the ideal Markovian squeezed-thermal limit.  For a moderate cascade of $N=5$ cavities with intrinsic loss ratio $\kappa_I/\kappa = 0.1$, the required source bandwidth is reduced to a quarter and a third for near-resonant and far-detuned targets, respectively. These results establish cascaded micro-rings, a structure already available on mature integrated-photonic platforms \cite{Alexander2025}, as a concrete, scalable route to generating the broadband squeezed reservoirs essential for quantum architectures.

%%%%%%%%%%%%%%%%%%%%%%%%%%%%%%%%%

The paper is organized as follows. We develops the model of the cascaded ring chain and the corresponding Heisenberg–Langevin equations in Section II, and discuss the resulting squeezing spectrum in Section III, including the uniform pumping case and a realistic regimes of intrinsic loss, non-uniform cavity frequencies, and pump attenuation. We then analyzes the system's robustness to manufacturing imperfections. In Section IV, we shows the convergence of flat spectrum to a Markovian reservoir that boost the light-matter coupling of an arbitrary target system. Finally, we conclude in Section V.

\section{Theoretical Model}
\begin{figure}
    \includegraphics[width= \columnwidth]{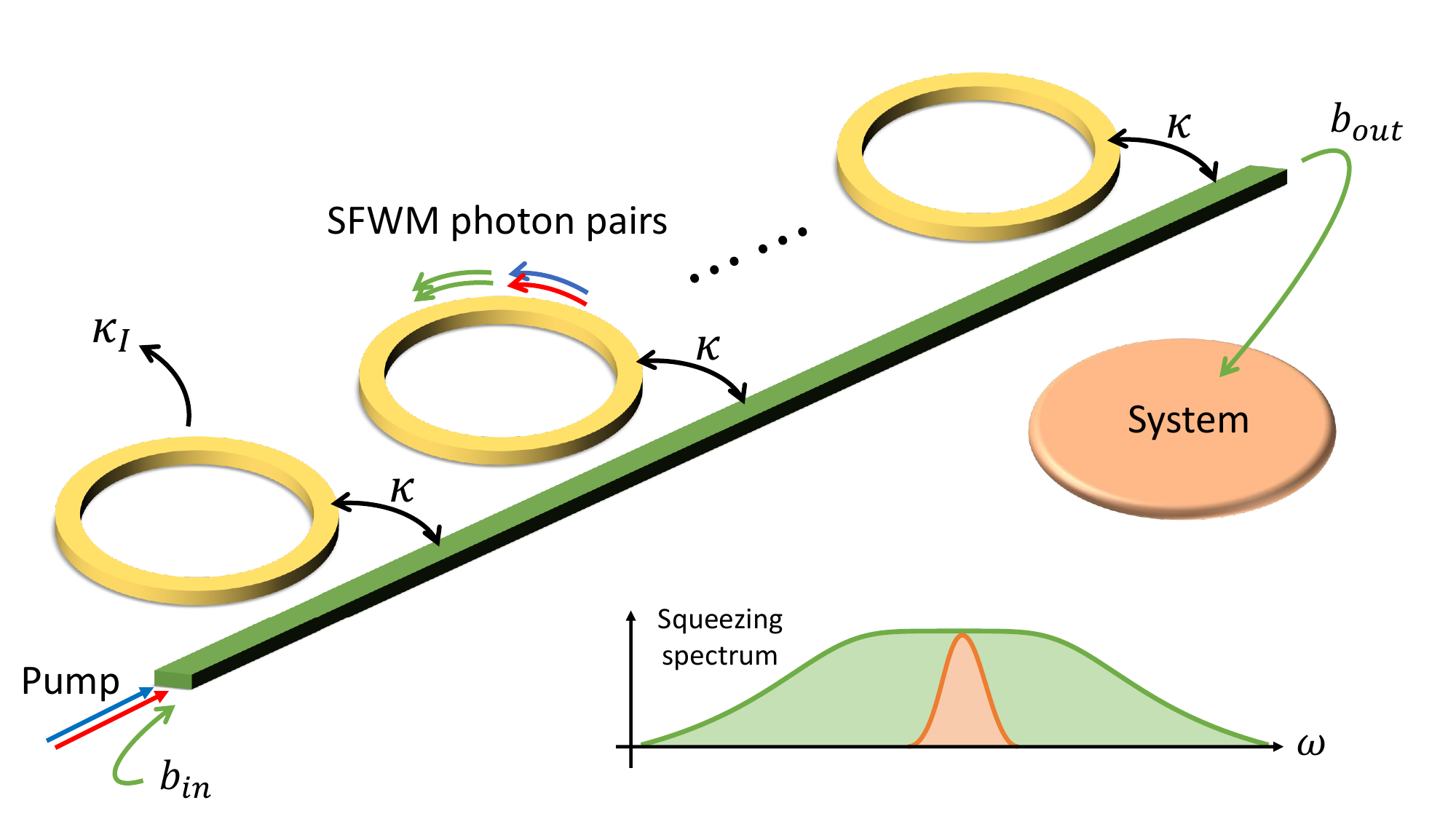}
    \caption{Schematic plot of the cascaded ring-resonator architecture. $N$ microrings is coupled to a common waveguide at a uniform rate $\kappa$, and to the environment at a rate $\kappa_I$. Spontaneous four-wave mixing creates degenerate photon pairs within each ring, denoted as parametric interaction $\frac{i\xi_k}{2} (a^{\dagger2}_k -a^2_k)$ for the $k$-th cavity. The broadband squeezed output can be routed to an arbitrary target system. When the bandwidth of the output $X$ quadrature bandwidth exceeds the target's characteristic bandwidth, the source effectively works as a squeezed reservoir that enhances the light-matter couplings. }
    \label{fig:1}
\end{figure}

We model the structure as a chain of $N$ microring cavities coupled in series along a bus waveguide, with input optical mode $b_{in}$, as shown in Fig. \ref{fig:1}. Each ring supports a single mode $a_k$ ($k = 1,\dots,N$) and is parametrically pumped to drive degenerate photon pairs at a rate $\xi_k$, with a detuning $\delta_k$ from half the sum of the two pump frequencies. Each mode decays into the bus at a rate $\kappa_k$ and into an intrinsic-loss reservoir at a rate $\kappa_I,_k$. The SLH formalism \cite{GoughJames2009,combes2017slh} yields an effective open-quantum-system description governed by the total Hamiltonian $H_{tot}$ and the loss operator $L_{tot}$

\begin{eqnarray}
H_{tot} = &\sum_{k=1}^N \left[-\frac{i\xi_k}{2}(a_k^{\dagger2} -a_k^2) + \delta_k a_k^\dagger a_k\right]  \nonumber \\ &+ \frac{1}{2i} \sum_{j>k=1}^N  \sqrt{\kappa_k \kappa_j} (a_j^\dagger a_k - a_k^\dagger a_j),
\end{eqnarray}
\begin{equation}
L_{tot} = \sum_{k=1}^N \sqrt{\kappa_k} a_k.
\end{equation}

The output field is determined by the standard input-output relation $b_{out} = b_{in} + L_{tot}$. Consequently, the Heisenberg-Langevin equation for the $k$-th ring follows directly

\begin{eqnarray}
\dot{a}_k &= -\left(i\delta_k + \frac{\kappa_k + \kappa_{I,k}}{2}\right) a_k - \xi_k a_k^\dagger \nonumber \\ &- \sum_{j=1}^{k-1} \sqrt{\kappa_k \kappa_j} a_j - \sqrt{\kappa_k} b_{in} - \sqrt{\kappa_{I,k}} c_{in, k},
\end{eqnarray}
where $b_{in}$ and $c_{in,k}$ represent the vacuum noise from the bus-mode and the intrinsic-loss environment coupled to the $k$-th cavity, respectively.

The parametric drive linearly mixes the two quadratures $X_k \equiv a_k + a_k^\dagger$ and $Y_k \equiv -i(a_k - a_k^\dagger)$ of each ring. Collecting these into the local quadrature vector $\mathbf{v}_k = (X_k,Y_k)^T$, with analogous definitions for $\mathbf{v}_{b}$ and $\mathbf{v}_{c,k}$, and applying a Fourier transform, we obtain the $N$ coupled Heisenberg-Langevin equations in the frequency domain,
\begin{equation}
-i\omega \mathbf{v}_k = -\mathbf{M}_k \mathbf{v}_k - \kappa \sum_{j=1}^{k-1} \mathbf{v}_j - \sqrt{\kappa} \mathbf{v}_{b} - \sqrt{\kappa_I} \mathbf{v}_{c,k},
\end{equation}
where the single-ring drift matrix is given by

\begin{equation}
\mathbf{M}_k = \begin{pmatrix} \Gamma_k + \xi_k & -\delta_k \\ \delta_k & \Gamma_k - \xi_k \end{pmatrix} \ ; \ \Gamma \equiv \frac{\kappa_k + \kappa_{I,k}}{2}.\end{equation}

Defining the single-ring susceptibility and transfer matrices as $\boldsymbol{\chi}_k(\omega) = (\mathbf{M}_k - i\omega \mathbf{I})^{-1}$ and $\mathbf{T}_k(\omega) \equiv \mathbf{I} - \kappa_k \boldsymbol{\chi}_k(\omega)$, respectively, we can express the cascaded output in terms of the input fields as
\begin{equation}
    \mathbf{v}_{out} = \mathbf{T}_{tot}(\omega) \mathbf{v}_{in} - \sqrt{\kappa_k \kappa_{I,k}} \sum_{k=1}^N \mathbf{C}_k(\omega) \mathbf{v}_{c,k},
\end{equation}
where $\mathbf{T}_{tot} = \mathbf{T}_N \mathbf{T}_{N-1} \cdots \mathbf{T}_1$ and $\mathbf{C}_k = \mathbf{T}_N \mathbf{T}_{N-1} \cdots \mathbf{T}_{k+1} \boldsymbol{\chi}_k$.

The squeezing properties are encoded in the symmetric output spectral matrix $\mathbf{S}_{out} \equiv \langle  \mathbf{v}_{out} \mathbf{v}_{out}^\dagger\rangle_{sym} = \frac{1}{2}\langle  \mathbf{v}_{out} \mathbf{v}_{out}^\dagger + (\mathbf{v}_{out} \mathbf{v}_{out}^\dagger)^T\rangle$. Because $\mathbf{v}_{in}$ and $\mathbf{v}_{c,k}$ are independent vacuum fields, we have the covariance matrices $\langle \mathbf{v}_{in} \mathbf{v}_{in}^\dagger\rangle_{sym} = I$ and $\langle \mathbf{v}_{c,i} \mathbf{v}_{c,j}^\dagger\rangle_{sym} = \delta_{ij}I$, with all other combinations vanishing. This yields the squeezing spectrum
\begin{equation}
\mathbf{S}_{out}(\omega) = \left[\mathbf{T}_{tot}(\omega) \mathbf{T}_{tot}^\dagger(\omega) + \sum_{k=1}^N \kappa_k \kappa_{I,k} \mathbf{C}_k(\omega) \mathbf{C}_k^\dagger(\omega)\right]_{sym}. \label{eq:fullspectrum}
\end{equation}
The first term is the input vacuum transmitted through the cascade, which is parametrically amplified along one quadrature and de-amplified along the other as it propagates through the cascade. The second term is the additive contribution from the $N$ uncorrelated intrinsic-loss vacuum.

To utilize the squeezed output from this cavity chain, it is directed into an arbitrary target system. The full $2\times 2$ matrix $\mathbf{S}_{out}$ contains more information than it is operationally needed. In standard cavity-QED setups, the output drives a localized device via a linear dipole interaction of the form $H_{int} \propto (a+a^\dagger)(b_{in} + b_{in}^\dagger) \propto X_a X_{in}$.  For couplings of this class, it has been shown \cite{c91x-bhqw,PRXQuantum.4.030316,PhysRevLett.120.030402}  that the induced dynamics depend exclusively on the two-time correlation of the driving field, $\langle X_{in}(t') X_{in}(t)\rangle$, for an initial gaussian state. These correlations are entirely captured by the $XX$ component of the frequency spectrum, $[\mathbf{S}_{out}(\omega)]_{XX}$, and are not dependent on the $Y$ quadratures and the off-diagonal correlations. Consequently, we will focus on the $XX$ spectrum, which is sufficient to fully characterize the influence of the cascaded source on downstream systems.

\section{Broadband squeezing spectrum}

In this section we evaluate the squeezing spectrum derived in Sec. II.  We begin with the simplest scenario, where the pump strength is uniform ($\xi _k = \xi$) and all cavities are resonant ($\delta_k = 0$). We also assume a uniform coupling ($\kappa_k = \kappa$) and loss rate ($\kappa_{I,k} = \kappa_I$) thorough out the rest of the paper. Under these ideal conditions, the equations of motion for the $X$ and $Y$ quadratures decouple, and the $XX$ component of the spectrum reduce to a more clear form
\begin{equation}
    [\mathbf{S}_{out}]_{XX} = |\beta(\omega)|^{2N} + \kappa \kappa_I |\chi(\omega)|^2 \left( \frac{1 - |\beta(\omega)|^{2N}}{1 - |\beta(\omega)|^2} \right), \label{eq:rspectrum}
\end{equation}
where $\beta(\omega) \equiv 1 - \kappa \chi(\omega) = 1 - \frac{\kappa}{\Gamma  + \xi - i\omega}$. The first term in Eq.\eqref{eq:rspectrum} describes the parametric attenuation of the input vacuum after $N$ passes through the squeezing medium. The second term of Eq.\eqref{eq:rspectrum} describes a geometric series accounting for the intrinsic-loss vacuum injected at each ring, which is subsequently attenuated by the downstream of the cascade. The squeezed-quadrature spectrum and the complete Gaussian field have different large-cascade behaviors. The X-quadrature spectrum pointwise converges to a constant limit $\frac{\kappa_I}{\kappa_I + 2 \xi}$ under $\xi<\Gamma/2$, the single ring stability condition. The anti-squeezed quadrature has a finite large-$N$ limit only when $2\xi <\kappa_I$. Figure 2(a) shows the spectrum for a finite series. For a single ring ($N=1$), the spectrum reduces to a standard Lorentzian profile; as the number of rings increases, the squeezing dip both broadens and deepens. While the transmission of input vacuum  $\sim|\beta(\omega)|^{2N}$ decays geometrically with $N$, the additive noise contribution, the second term in Eq.\eqref{eq:rspectrum}, saturates. Because $|\beta(\omega)| < 1$, the asymptotic limit yields a squeezing plateau whose ultimate depth is determined entirely by the ratio of intrinsic loss to bus coupling.

\begin{figure}
    \includegraphics[width= \columnwidth]{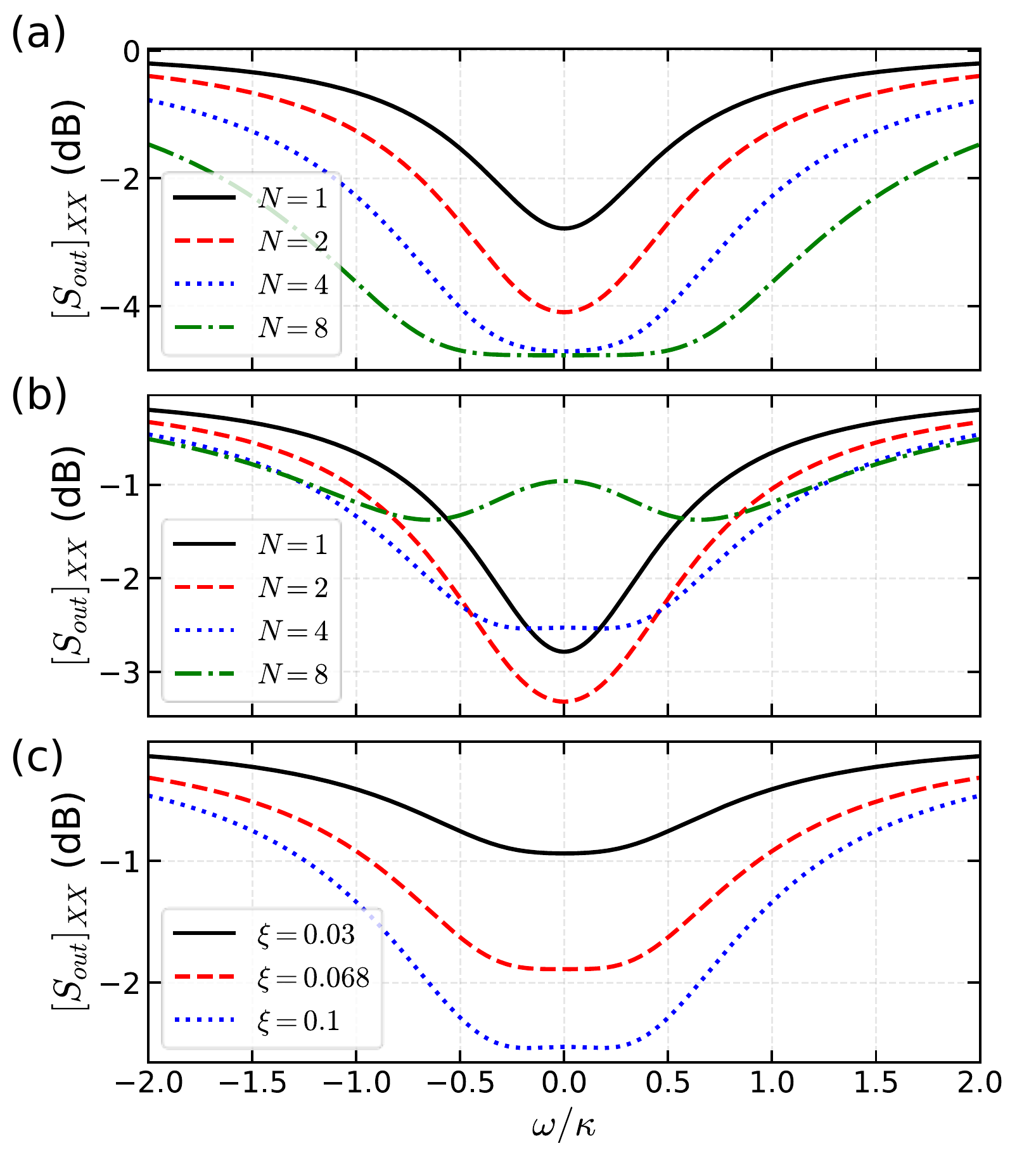}
    \caption{Output $X$-quadrature squeezing spectra for a cascade of resonant microrings ($\delta_k = 0$) with uniform intrinsic loss $\kappa_I = 0.1\kappa$. (a) Uniform pump strength ($\xi_k = \xi = 0.1\kappa$) evaluated across various cascade lengths $N$. As $N$ increases, the cumulative parametric interaction forms a flat, asymptotic squeezing plateau. (b) Spectra under realistic geometric pump attenuation ($\xi_k = r^{k-1}\xi$) for various $N$, utilizing the same initial pump strength $\xi = 0.1\kappa$. The spatially varying pump profile degrades the plateau into a sharp central dip at $\omega=0$. (c) Optimization of the depleted-pump cascade for a fixed length of $N=4$ under various initial pump strengths $\xi$. By tuning the pump near the zero-curvature condition, $\xi \approx 0.068\kappa$, a broad, locally flat squeezing profile is recovered.}
    \label{fig:2}
\end{figure}

While a uniform pair-generation rate requires significant effort to engineer \cite{zhao2018topological,doi:10.1126/science.abg3904}, it serves as an baseline for its perfectly flat, broadband nature. In a practical device, the pump field propagates along the bus waveguide and depletes as it drives the cascaded cavities. Let $\alpha_{in,a}^{(k)}$ and $\alpha_{k,a}$ denote the input and intracavity fields of the $k$-th ring, respectively, where $a \in \{s,i\}$ denotes the signal and idler fields. Assuming the same loss rates as pump modes, the steady-state dynamics are governed by the semiclassical equation of motion $\dot{\alpha}_{k,a} = - \Gamma \alpha_{k,a} - \sqrt{\kappa}\alpha_{in,a}^{(k)}$ and the iterative input-output relation $\alpha_{in,a}^{(k+1)} = \alpha_{out,a}^{(k)} = \alpha_{in,a}^{(k)} + \sqrt{\kappa} \alpha_{k,a}$. Solving for the steady state and noting that the spontaneous four-wave mixing (SFWM) pair-generation rate scales as $\xi_k \propto |\alpha_{k,s}\alpha_{k,i}|$, we find that the pump strength depletes geometrically
\begin{equation}
    \xi_k  = \left(\frac{\kappa-\kappa_I}{\kappa + \kappa_I}\right)^{2k-2} \xi, \label{geometric}
\end{equation}
where $\xi$ is the generation rate in the first ring. Figure 2(b) illustrates the spectrum when imposing this geometric attenuation for the same initial pump strength used in Fig. 2(a). The attenuation degrades the squeezing plateau at large $N$. 

Two competing distortions shape this spectrum relative to the uniform-pump result. On the one hand, each individual ring contributes a Lorentzian dip whose depth and width are dictated by $\xi_k$. The early rings in the chain ($k$ small, $\xi_k$ large) produce deep, broad notches near $\omega=0$, while the late rings ($k$ large, $\xi_k$ small) contribute only weak, narrow attenuation. The cumulative product compounds all the contributions. Consequently, the cascade tends to develop a central dip within a dip, a sharply peaked minimum at $\omega=0$ rather than a flat plateau. Because a Markovian squeezed reservoir needs the absence of such a feature, it is critical to identify the parameter regime where these two distortions cancel and the spectrum remains locally flat near the carrier frequency. This can be done by considering a vanishing spectral curvature at resonance, which is equivalent to $\frac{d^2}{d\omega^2} [\mathbf{S}_{out}(\omega=0)]_{XX} = 0$ because of the symmetry. Vanishing curvature at the carrier removes the leading quadratic variation of the quadrature spectrum, but it does not by itself specify a broadband interval. We use this condition as an analytic design rule determining the existence of nearly flat curve. Imposing the zero-curvature condition generates a polynomial that does not factorize into a simple form. However, by analyzing its behavior in the weak driving $\xi \to 0$ limit, we can identify the maximum cascade length $N$ that possesses a stable, spectrum-flattening pump strength $\xi$,
\begin{equation}
 N \lesssim \frac{1}{2} \left( \frac{\kappa_I}{\kappa} +  \frac{\kappa}{\kappa_I}\right).
\end{equation}
We note that reversing the sign of $\xi$ yields an anti-squeezed spectrum, which also reverses the flatness condition. Maintaining a flat profile in the anti-squeezed quadrature requires a larger number of rings for a given loss ratio $\kappa_I/\kappa$.

Figure 2(c) demonstrates a specific configuration ($N=4$, $\kappa_I/\kappa = 0.1$) where the spectrum is successfully flattened, exhibiting both greater bandwidth and stronger squeezing than a single ring without additional pump power. For the exact zero-curvature strength of $\xi \approx 0.068\kappa$, the curve has a constant plateau of range around $0.7\kappa$. Even if the pump strength is increased to $\xi = 0.1\kappa$, the spectrum maintains satisfactorily flat for a range around $0.9 \kappa$ while further enhancing the overall squeezing.

%%%%%%%%

\begin{figure}
    \includegraphics[width= \columnwidth]{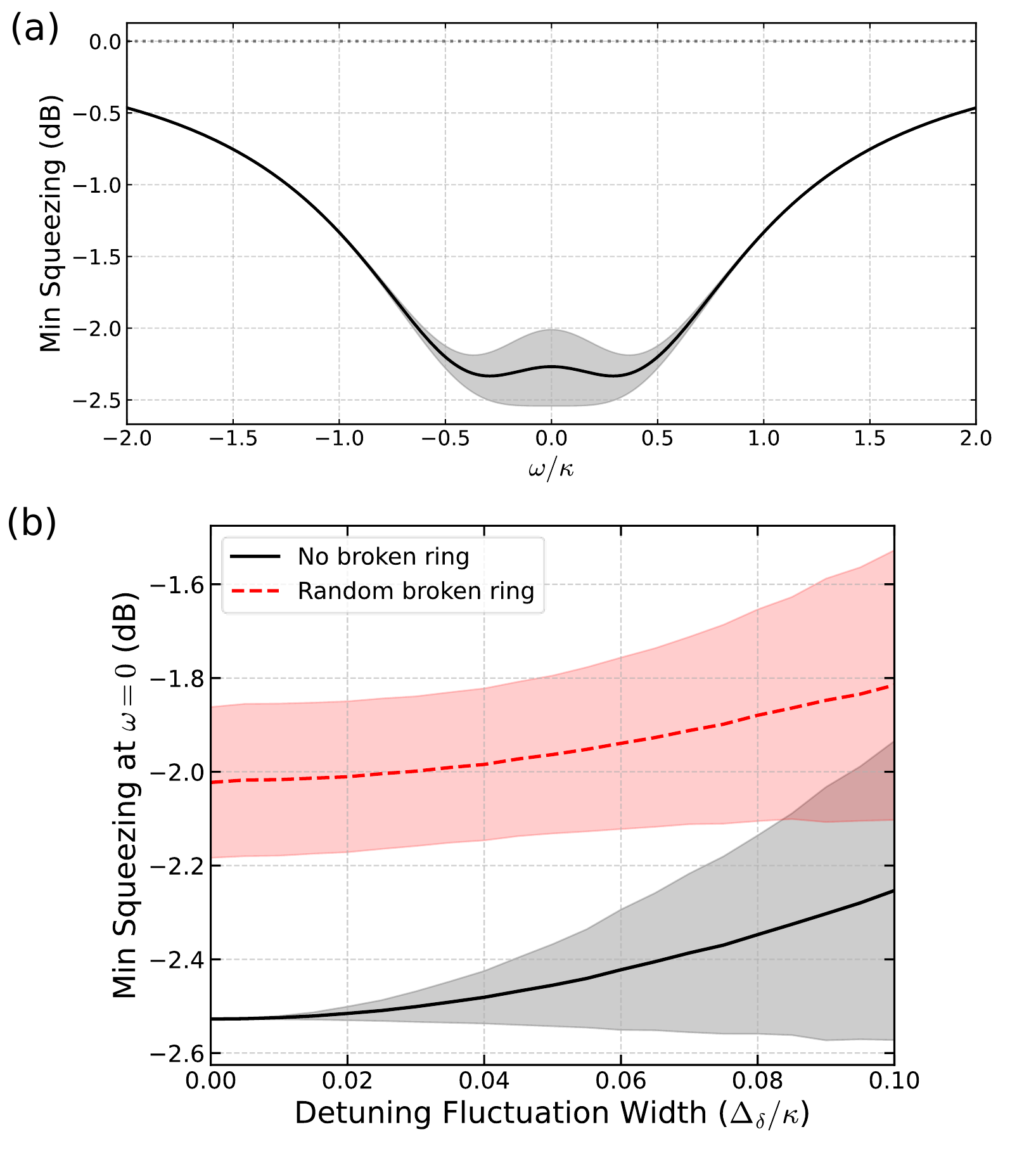}
    \caption{Cascaded source against fabrication disorder and broken ring. (a) Average and standard deviation of the principal squeezed spectrum for $N=4$, $\xi= \kappa_I=0.1\kappa$, evaluated over an ensemble of Gaussian random frequency shifts with a standard deviation of $\Delta_\delta = 0.1\kappa$.
    Although the disorder impacts the squeezing depth near resonance, the overall flatness of the broadband plateau is preserved. (b) Resonant squeezing magnitude (average and standard deviation) as a function of the frequency fluctuation width $\Delta_\delta$. All other parameters are the same. We also compare the performance under severe defect scenario in which a randomly selected ring is broken. Even under the combined degradation of a ring failure and a $\Delta_\delta = 0.1\kappa$ frequency variance, the total squeezing penalty is restricted to around $0.2$ dB relative to an $-1.8$ dB average.}
    \label{fig:3}
\end{figure}

Next, we investigate the robustness of the microring array against fabrication imperfections, specifically detuned and defective cavities. Although active control techniques, such as electro-optic \cite{dutt2019experimental,Zeng2026,Li2021,balcytis2022synthetic,PhysRevLett.132.123802,10.1063/5.0056359} or thermo-optic \cite{Chen2025,Zhang:19} modulation, are employed to dynamically tune resonance frequencies, evaluating the tolerance of the uncompensated array is still crucial due to thermal effects induced by pumping in practical experiments. We revisit the full spectral matrix in Eq. \eqref{eq:fullspectrum}, introducing a random frequency shift $\delta_k$ for each ring. Nonzero shifts couple the $X$ and $Y$ quadratures, rotating the principal axis of maximum squeezing. This rotation can be compensated by adjusting the relative phase of the parametric pump. Fig. 3(a) displays the average and standard deviation of the principal squeezed spectrum under Gaussian distributed random cavity frequencies with a standard deviation of $\Delta_\delta = 0.1\kappa$, evaluated under the geometric pump attenuation condition Eq. \eqref{geometric}. For this degree of disorder, the fluctuation in the squeezing depth is confined to approximately $0.2$ dB, safely preserving the local flatness of the spectrum. Figure~3(b) illustrates the growth of this standard deviation as a function of the disorder width $\Delta_\delta$. We confirm that fluctuations vanish in the ideal limit $\Delta_\delta \to 0$. Furthermore, we consider a severe defect scenario in which a randomly selected ring fails to generate photon pairs ($\xi_k = 0$) but still deplete the pump field, thereby preserving the geometric attenuation sequence. The red dashed line in Fig. \ref{fig:3}(b) shows the result of randomly disabling one of the four rings. While this defect degrades the baseline resonant squeezing, the combined effect of a broken ring and a $\Delta_\delta = 0.1\kappa$ resonance frequency fluctuation induces a penalty of only $\sim0.2$ dB relative to an average $-1.6$ dB squeezing depth. In practical implementations, designing the cavities with a relatively low quality factor ($Q$) ensures that the squeezed output fully envelops the resonance of the downstream target. The small standard deviation induced by these disorders guarantee that the spectrum remains reliably flat across this broad regime. Because both the squeezing magnitude and the optimal quadrature angle vary slowly across the characteristic linewidth of the target, the pump phase can simply be tuned to align the optimal squeezing axis with the frequency of the target system.

\section{Finite-bandwidth bath flatness and target response}

The plateau condition of Sec. III guarantees that a properly tuned cascade produces a broad, flat squeezing spectrum. This spectral flatness is a needed source to function as the Markovian squeezed reservoir, which requires the reservoir's time-domain correlation function to approximate a memoryless Dirac delta pulse. In this section we make this requirement quantitative. By characterizing the non-Markovianity of the induced system dynamics, we demonstrate that the cascade's flat-topped spectrum approaches the ideal Markovian limit faster than a single-resonator Lorentzian of equal squeezing depth.

As discussed in Sec. I, when the chain output drives a linearly coupled target, the reduced dynamics of the target depend on the two-time correlation function of the input field, $\langle X_{in}(t') X_{in}(t)\rangle$. The Fourier transform of this correlation function is exactly the $XX$ spectral component $[\mathbf{S}_{out}(\omega)]_{XX}$ given in Eq. \eqref{eq:fullspectrum}. As detailed in Appendix A, the frequency-domain response $R(\omega)$ of a target system with resonance frequency $\delta$ and linewidth $\kappa_s$ to this broadband input is given by the product of the target's Lorentzian susceptibility and the source's squeezing spectrum
\begin{equation}
    R(\omega) = \frac{\kappa_s}{(\omega-\delta)^2 + \kappa_s^2/4} [\mathbf{S}_{out}(\omega)]_{XX}.
\end{equation}

To compare the performance of different spectral shapes, we separate the reservoir-induced response into a memoryless component and a memory-bearing component. We decompose the spectrum as $[\mathbf{S}_{out}(\omega)]_{XX} = S^{ml} + S^{mem}(\omega)$, where the memoryless baseline is defined by the squeezing depth at the target's resonance, $S^{ml} \equiv [\mathbf{S}_{out}(\delta)]_{XX}$. Because the response function is linear, this induces a corresponding split $R(\omega) = R^{ml}(\omega) + R^{mem}(\omega)$.  Following the measure of non-Markovianity in Ref. \cite{PhysRevA.92.062306} , which is defined through the difference between an exact system two-time correlation function and its Markov-limit counterpart, we further quantify the relative deviation from the ideal Markovian limit using the $\mathcal{L}^1$ norm ratio of total responds in the time domain
\begin{equation}
    \mathcal{R}_N \equiv \frac{\|R^{mem}(t)\|_1}{\|R^{ml}(t)\|_1} = \frac{\int dt |R^{mem}(t)|}{\int dt|R^{ml}(t)|},
\end{equation}
where $R^{ml}(t)$ and $R^{mem}(t)$ are the inverse Fourier transforms of their respective frequency-domain responses. This quantity measures the target-filtered response departure of the $X$-quadrature spectrum from a constant spectrum approximation. By this definition, the cascaded source approaches to an ideal Markovian squeezed reservoir when $\mathcal{R}_N \ll 1$. This limit is achieved when the memory-bearing fluctuation $S^{mem}(\omega)$ is negligibly small across the span of the target's Lorentzian profile.

\begin{figure}
    \includegraphics[width= \columnwidth]{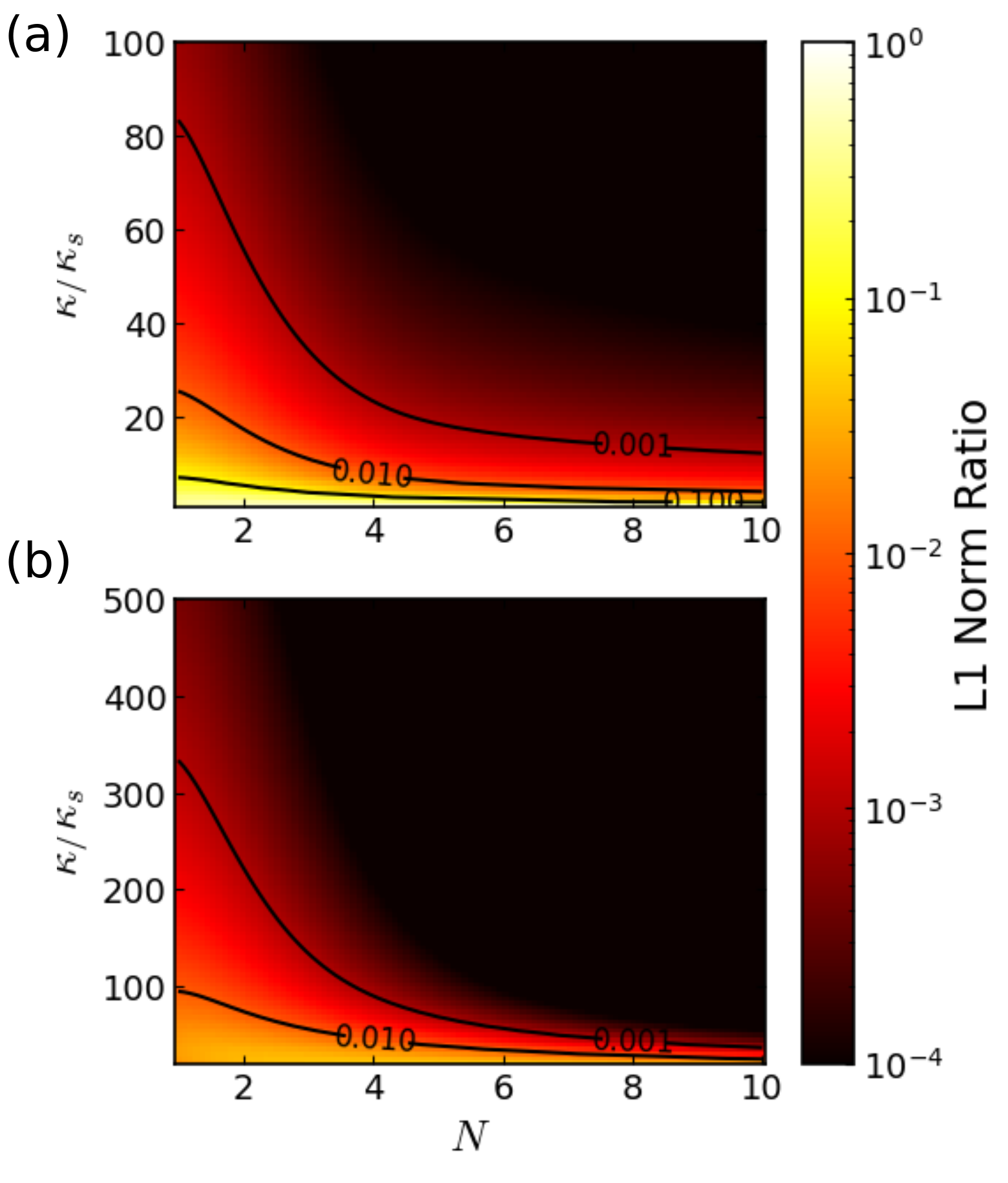}
    \caption{Color map of the ratio $\mathcal{R}_N$ as a function of the cascade length $N$ and the source-to-target linewidth ratio $\kappa/\kappa_s$. The reservoir performance is evaluated for a target (a) near resonance $\delta = \kappa_s$ and at (b) far detuning $\delta = 20\kappa_s$. The ratio of intrinsic loss and pump strength are fixed at $\kappa_I/\kappa = 0.2$ and $\xi/\kappa = 0.1$, respectively. The spectrum is analytically continued to continuous $N$ to smoothly visualize the parameter scaling. The significant reduction in the required linewidth ratio $\kappa/\kappa_s$ at larger $N$ illustrates how the flat-topped cascaded cavities relaxes the extreme bandwidth constraints needed to reach an ideal Markovian response.}
    \label{fig:4}
\end{figure}

Figure.\ref{fig:4} illustrates the norm ratio $\mathcal{R}_N$ for a resonant chain driven by a uniform pump, which suffices to demonstrate the advantage of spectral flatness over squeezing depth alone. To visualize the scaling trends smoothly, we analytically continue Eq. \eqref{eq:rspectrum} to evaluate $N$ as a continuous real parameter, providing a clear comparison with the single-resonator ($N=1$) baseline. We plot $\mathcal{R}_N$ as a function of the source-to-target linewidth ratio, $\kappa/\kappa_s$, to determine the spectral scaling required to effectively approximate an ideal Markovian reservoir. We scale the source linewidth $\kappa$ while holding the ratios $\kappa_I/\kappa$ and $\xi/\kappa$ constant, ensuring that the resonant squeezing depth remains invariant for fixed $N$.

Two cases are examined: a near-resonant target ($\delta = 1\kappa_s$) in Fig. \ref{fig:4}(a) and a far-detuned target ($\delta = 20\kappa_s$) in Fig. \ref{fig:4}(b). In both scenarios, the flat-topped spectrum of the longer cascade shows a advantage over the single resonator, reducing the source bandwidth required to achieve the same degree of Markovianity. For a near-resonant target, reaching a stringent non-Markovian limit of $\mathcal{R}_N = 10^{-3}$ with a single ring requires a linewidth ratio of $\kappa/\kappa_s \approx 80$. For a five-ring $N=5$ chain, this requirement is cut to a quarter to $\kappa/\kappa_s \approx 20$. This advantage is still pronounced for the far-detuned target, where the requisite linewidth ratio for the same $\mathcal{R}_N$ drops from 300 at $N=1$ down to 100 at $N=5$. Because the single-ring Lorentzian spectrum rolls off slowly as $\omega^{-2}$ beyond its linewidth, its memory-bearing component $S^{mem}(\omega)$ retains appreciable weight even at moderate detunings. Consequently, its norm ratio $\mathcal{R}_N$ decreases only slowly as the source is broadened.

This rapid convergence to the Markovian limit have a experimental advantage regarding the required resonator quality factor ($Q$). For a single Lorentzian source, achieving high Markovianity demands a massive linewidth $\kappa$ (low $Q$) and a proportionally large parametric gain $\xi$ to maintain a fixed squeezing depth. These simultaneous requirements for heavy coupling and intense pumping severely constrain practical device design. The cascaded architecture relaxes these requirements. Because the flat-topped spectrum achieves equivalent Markovianity at a significantly narrower per-ring linewidth, each resonator can operate at a higher individual $Q$. Consequently, the deep squeezing normally demanded from a single, strongly pumped stage is instead achieved by $N$ modestly pumped rings. This approach replaces the extreme low-$Q$/high-gain constraints of conventional broadband sources with milder per-ring specifications, suited to the loss budgets of modern integrated photonic platforms.
%%%%%%

\section{Conclusion}
In conclusion, we have proposed a scheme in which a cascaded chain of parametric cavities generates a broadband squeezed vacuum that approximates a stationary Markovian squeezed reservoir. Our central result is that the flat-topped spectrum of the cascade converges to the ideal Markovian limit significantly faster than the Lorentzian spectrum of a single resonator of equal squeezing depth. We find the conditions required to generate such a spectrum using a specific number of rings. The advantage of this multi-ring structure is that it relaxes the demanding low-$Q$ and high-gain requirements of a single broadband cavity, distributing the necessary squeezing depth across a series of moderately pumped cavities.

We emphasize that our analysis is not restricted to ring resonators. The underlying theoretical framework extends naturally to any cascaded physical system capable of parametric interactions, including bulk optical parametric oscillators \cite{Takanashi:19,PhysRevResearch.7.023110,PhysRevLett.124.171102}, superconducting parametric cavities \cite{Grimm2020,iyama2024observation,Zhong_2013,PhysRevLett.128.153603}, and optical parametric amplifiers \cite{10.1063/5.0144385, Jankowski:22}. Regarding on-chip implementations, the building blocks, parametrically driven micro-rings, have already been demonstrated beyond silicon nitride across several integrated platforms, including lithium niobate \cite{ren2026,doi:10.1126/sciadv.aeb5758,Wang:24,Lu:19} and silicon carbide \cite{guidry2021high,Guidry:20,Fan:18}. Furthermore, the multi-ring structures, such as coupled-resonator optical waveguides (CROWs), have been theorized \cite{Yariv:99,Poon:04,PhysRevB.96.184304} and experimentally characterized \cite{Cooper:10,Jayatilleka:19} already. The demonstrated per-ring squeezing and mature cascaded-resonator fabrication confirm our scheme within the reach of current technology. We anticipate that this straightforward design can be adopted through diverse platforms, providing a pathway to engineer a squeezing reservoir that benefits hardware in quantum technologies.

\section*   {ACKNOWLEDGMENTS}
We acknowledge Allen Zhang for insightful discussions and helpful feedback that substantially improved this work.

\appendix
\section{Derivation of the input response of arbitrary systems}
To evaluate the non-Markovianity of the target system, we derive its two-time correlation function in response to the driving input field. Consider a target system with Hamiltonian $H_{tar} = H_a + H_{int} + H_{res}$ coupled linearly to a waveguide, where
\begin{equation}
    H_a = \omega_a a^\dagger a,
\end{equation}
\begin{equation}
    H_{int} = \sqrt{\frac{\kappa_s}{2\pi}} \int d\omega \, (a+a^\dagger)(b_\omega + b_\omega^\dagger).
\end{equation}
Here, $a$ is the annihilation operator for the target cavity mode, $b_\omega$ is the continuous waveguide mode at frequency $\omega$, and $H_{res}$ is the remaining part unrelated to $b_\omega$. 

The Heisenberg-Langevin equation for $a$ is driven by both the target quadrature $X_a \equiv a + a^\dagger$ and the waveguide input quadrature $X_{in}(t) \equiv b_{in}(t) + b_{in}^\dagger(t)$, where the temporal input operator is defined as $b_{in}(t) \equiv \frac{1}{\sqrt{2\pi}} \int d\omega \, b_\omega e^{i \omega t}$
\begin{equation}
    \dot{a}(t) = -i\omega_a a(t) - \frac{\kappa_s}{2}X_a - i\sqrt{\kappa_s} X_{in}(t) - i[a(t), H_{res}].
\end{equation}
Neglecting the internal dynamics $H_{res}$ to isolate the waveguide response, the formal solution can be expressed in vector form as
\begin{equation}
    \mathbf{v}(t) = e^{\mathbf{M}(t-t_0)}\mathbf{v}(t_0) + \int_{t_0}^t ds \, e^{\mathbf{M}(t-s)} X_{in}(s) \boldsymbol{\eta},
\end{equation}
where we define
\begin{equation}
    \mathbf{v}(t) \equiv \begin{pmatrix} a(t) \\ a^\dagger(t) \end{pmatrix}, \quad \boldsymbol{\eta} \equiv -i\sqrt{\kappa_s} \begin{pmatrix} 1 \\ -1 \end{pmatrix},
\end{equation}
\begin{equation}
    \mathbf{M} \equiv \begin{pmatrix} -i\omega_a - \frac{\kappa_s}{2} & -\frac{\kappa_s}{2} \\ -\frac{\kappa_s}{2} & i\omega_a - \frac{\kappa_s}{2} \end{pmatrix}.
\end{equation}

The matrix exponential is solvable to
\begin{equation}
    e^{\mathbf{M}t} = e^{-\kappa_s t/2} \left[\cos(\Omega t) \mathbf{I} + \frac{\sin(\Omega t)}{\Omega} \left(\mathbf{M} + \frac{\kappa_s}{2}\mathbf{I}\right)\right],
\end{equation}
where $\Omega \equiv \sqrt{\omega_a^2 - \kappa_s^2/4}$. In the rotating wave limit $\omega_a \gg \kappa_s$, we can approximate $\Omega \approx \omega_a$. This approximation decouples the evolution of $a$ and $a^\dagger$, while still preserving the rapidly oscillating components in $X_{in}$. 

Taking the long-time limit ($t_0 \to -\infty$) to eliminate initial-state dependencies, the steady-state waveguide contribution to the target's two-time correlation function is
\begin{eqnarray}
    \langle a^\dagger(t+\tau) a(t) \rangle = \kappa_s \int_{0}^{\infty} du' \int_{0}^{\infty} du \, e^{(-\kappa_s/2+i\omega_a)u'} \nonumber \\ \times e^{(-\kappa_s/2-i\omega_a)u} \langle X_{in}(t+\tau-u) X_{in}(t-u') \rangle.
\end{eqnarray}
Since we assume the state reaches stationary, the time-translation symmetry implies $\langle X_{in}(t+\tau-u) X_{in}(t-u') \rangle \equiv C(\tau-u+u')$ for some function $C$. Evaluating the integral, the correlation function simplifies to a convolution, $\langle a^\dagger(t+\tau) a(t) \rangle = \int_{-\infty}^{\infty} d\xi \, K(\xi) C(\tau-\xi)$, with the temporal kernel
\begin{equation}
    K(\xi) \equiv \frac{\kappa_s}{2} e^{-i\omega_a \xi} \int_{|\xi|}^\infty du \, e^{-\kappa_s u /2} = e^{-\kappa_s |\xi|/2 - i\omega_a \xi}.
\end{equation}
The Fourier transform of $K(\xi)$ is precisely a Lorentzian function centered at $\omega_a$ with linewidth $\kappa_s$. Consequently, the steady-state frequency response of the target system is simply the product of this target Lorentzian susceptibility and the Fourier transform of the input $X$-quadrature correlation function, which is the cascaded source's squeezing spectrum, $[\mathbf{S}_{out}(\omega)]_{XX}$.

\bibliography{CascadedMicroRing.bib}% Produces the bibliography via BibTeX.

\end{document}